# HABITABLE ZONES OF POST-MAIN SEQUENCE STARS

Ramses M. Ramirez[1,2] and Lisa Kaltenegger[1,3]

[1]Carl Sagan Institute, Cornell University, Ithaca, NY
[2]Cornell Center for Astrophysics and Planetary Science, Cornell University, Ithaca, NY
[3]Department of Astronomy, Cornell University, Ithaca, NY

**ABSTRACT**

Once a star leaves the main sequence and becomes a red giant, its Habitable Zone (HZ) moves outward, promoting detectable habitable conditions at larger orbital distances. We use a one-dimensional radiative-convective climate and stellar evolutionary models to calculate post-MS HZ distances for a grid of stars from 3,700K to 10,000K (~M1 to A5 stellar types) for different stellar metallicities. The post-MS HZ limits are comparable to the distances of known directly imaged planets. We model the stellar as well as planetary atmospheric mass loss during the Red Giant Branch (RGB) and Asymptotic Giant Branch (AGB) phases for super-Moons to super-Earths. A planet can stay between 200 million years up to 9 Gyr in the post-MS HZ for our hottest and coldest grid stars, respectively, assuming solar metallicity. These numbers increase for increased stellar metallicity. Total atmospheric erosion only occurs for planets in close-in orbits. The post-MS HZ orbital distances are within detection capabilities of direct imaging techniques.

*Key words:* planets and satellites: atmospheres, planets and satellites: detection, planet-star interactions, stars: post-main-sequence, stars: winds, outflows

## 1. Introduction

The recent discovery of sub-Earth sized planets around an 11 billion year old star (Campante et al. 2015) illustrates that planets have formed early in the history of the universe. Although none of the 5 planets found are located in the Habitable Zone, this discovery shows that such planets could be found in the near-future in the HZ of their star (e.g. in the Kepler dataset). This finding also opens the interesting question of where the Habitable Zone is located during the later stages of a stars' evolution. Kepler 444 is an 11 Gyr (+/-) main-sequence K0 star and is therefore still on the main-sequence, but a hotter star type would already have moved onto the giant star branch at that age. The stellar post-MS phase presents a serene environment that may promote habitability (Stern 2003). Here, we calculate the post-main-sequence boundaries of the Habitable Zone as well as discuss potential habitability for planets around such evolved stars.

The HZ is the circumstellar region in which liquid water could exist on the surface of a rocky planet. The HZ is a tool that guides missions and surveys in prioritizing planets for follow up observations. Although planets located outside the HZ are not excluded from hosting life, detecting biosignatures remotely on such planets should be extremely difficult. The empirical habitable zone boundaries used are those originally defined using a 1D climate model by Kasting et al. (1993), and updated for Main-Sequence (MS) stars with effective temperatures ($T_{eff}$) between 2,600 and 7,200K (Kopparapu et al. 2013;2014, Ramirez et al., 2014ab), including additional inner edge estimates based on 3D models (Leconte et al. 2013a). We expand the model to $T_{eff}$ up to 10,000 K (~A5 spectral class).

However, these former papers have focused on the Main Sequence HZ. Danchi and Lopez (2013) used a parameterization of the Main Sequence HZ limits (Kaltenegger & Sasselov 2011, Selsis et al. 2007, Kasting et al. 1993) from a previous version of the 1-D climate model for stars with $T_{eff}$ between 3,700 and 7,200K to estimate the HZ in a first approximation for post-MS stars for a subset of stellar masses. These authors showed how stellar metallicity impacts the duration that a planet at a certain orbital distance can spend in the Habitable Zone (Danchi and Lopez, 2013). Ramses and Kaltenegger (2014) have recently extended the HZ boundary calculations to the pre-MS using the updated 1D climate model (Ramirez et al., 2014ab), providing additional targets in the wider pre-MS HZ for future observations. Habitable conditions during the pre-main-sequence of a stars' lifetime may persist for nearly 2.5 Gyr for cool stars (Ramirez and Kaltenegger, 2014). In this paper we explore habitable conditions during the stellar post-main-sequence. We use this updated 1-D climate model to calculate the limits of the post-MS HZ with input from stellar evolutionary models to capture the changing stellar energy distribution (SED) and luminosity of evolved stars and their influence on the HZ limits (Bertelli et al., 2008;2009; Dotter et al., 2007;2008) for stars with masses from 0.5 to 1.9 Solar masses and corresponding spectral classes of A5 to M1.

Life may become remotely detectable during the post main sequence lifetime of a star. First, life may be able to evolve quickly (i.e. within a few million to a hundred million years). On Earth, isotopic data indicate that life on Earth started by about 3.8 Gya (Mojzsis & Arrhenius, 1997) but could have evolved even earlier but was destroyed during the period of late heavy bombardment (ibid). This inference is supported by the observation that the earliest signs of life on Earth coincide with the tail end of this heavy bombardment phase (ibid).

Secondly, it is not necessary for life to evolve during the post-MS phase. Life may have started in an initially habitable environment (e.g. during the star's pre-MS phase (Ramirez & Kaltenegger 2014) and then moved subsurface, or stayed dormant until surface conditions allowed for it to move to the planet's surface again, like in a stars' post-MS phase. Lastly, life could have evolved during early times on a cold planet located beyond the traditional habitable zone, remaining subsurface or under a layer of ice until emerging during the post-MS phase. In our own Solar System, if life exists in the subsurface ocean of icy exo-moons like Europa or Enceladus, this life may be exposed during our Sun's red giant branch phase (RGB), during which the post-MS HZ will move outward to Jupiter's orbit, allowing atmospheric biosignatures to potentially become remotely detectable at those orbital distances. For planets or moons as small as Europa, such atmospheric signatures would be short-lived due to the low gravity. But for super-Europa analogues or other habitable former icy planets such atmospheric signatures could build up. Higher disk densities around massive stars may translate into more massive objects than in our Kuiper belt region (~ 3 times the terrestrial planets; Gladman et al., 2001). Such planets may be present around current post-main-sequence stars.

We also assess the role that planetary *atmospheric erosion* during the post-main-sequence has on habitability. As the star expands through the Red Giant Branch (RGB) and Asymptotic Giant Branch (AGB), stellar mass losses produce high stellar winds, which can erode planetary atmospheres (e.g. Lorenz et al., 1997; Villaver and Livio, 2007), see section 3.3 and 3.4. In contrast, low EUV fluxes during the post-main-sequence limit EUV-driven hydrodynamic escape for planets in the post-MS HZ, making it only a secondary concern.

Our models are described in Section 2, post-MS HZ boundaries, atmospheric loss, time in the post-MS HZ as well as comparison of the post-MS HZ distance to detected exoplanets are given in Section 3. We discuss assumptions in Section 4, followed by concluding comments in section 5.

## 2. Methods

### 2.1 Post-Main Sequence stellar grid models

We compute the post-MS luminosities for a grid of stars from A5 to M1 (Fig. 1) using the corresponding stellar masses: $1.9 M_{Sun}$, $1.5 M_{Sun}$, $1.3 M_{Sun}$, $1 M_{Sun}$, $0.75 M_{Sun}$, and $0.5 M_{Sun}$ based on the Padova stellar evolutionary tracks for the high-mass stars (Bertelli et al., 2008; 2009) (1.9, 1.5, 1.3, and 1 Msun) and the corresponding Darmouth models for the lower mass stars in our grid (Dotter et al., 2007; 2008)(0.75, and $0.5 M_{Sun}$).All calculations start at the beginning of the red giant branch with stellar luminosities increasing until the tip of the RGB. For stars of solar mass and greater, these luminosities decrease coincident with the helium flash, before again increasing along the AGB (Fig. 1). For less massive stars, stellar winds during the RGB phase reduce their masses below ~0.5 that of the present Sun, too low to undergo the AGB phase (see e.g. Iben, 2013, p. 616).

The maximum stellar luminosities of the grid stars are between 40,000 and 1,000 times their values at the Zero Age Main Sequence (~1,000, 4500, 40,000 for the F1, Sun, and M1 star model, respectively, see Fig. 1). The changing stellar effective temperature, $T_{eff}$, and SED throughout the post-MS impacts the relative contribution of absorption and scattering of incoming light in the planetary atmosphere. That, in turn, changes the effective stellar flux at the top of the atmosphere of the planet, $S_{eff}$. Therefore, both $S_{eff}$ as well as the orbital distance of the post-MS HZ boundaries, change through time (see Figs. 2-3).

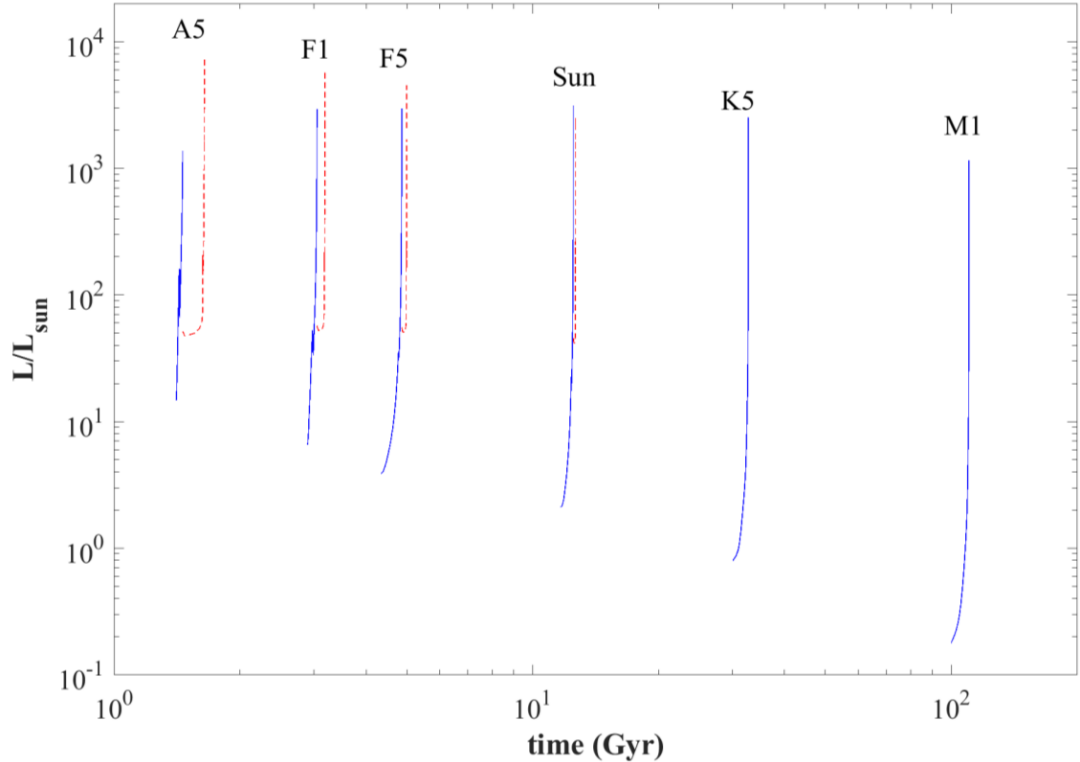

*Figure 1: Evolution of stellar luminosity for our grid of stars (A5, F1, F5, Sun, K5, M1) for the red giant branch (solid blue lines) and the horizontal plus asymptotic giant branch (red dashed lines).*

### 2.2 Post-Main Sequence Habitable Zone models

The inner edge of the empirical HZ (Kasting et al. 1993) is defined by the stellar flux received by Venus when we can exclude the possibility that it had standing water on the surface (about 1 Gyr ago), equivalent to a $S_{eff}$ value of 1.77. The outer edge is defined by the stellar flux that Mars received at the time that it may have had stable water on its surface (about 3.8 Gyr ago), which corresponds to $S_{eff} = 0.32$ in our Solar System. We also show (Fig. 2-3) a more conservative inner edge of the HZ (dashed lines) based on 3D models that calculate when a runaway greenhouse scenario would occur on a planet based on current Earth-based atmosphere and cloud models, equivalent to $S_{eff}$ ~1.11 in our own Solar System (Leconte et al. 2013a). Planets become too hot and get devolatilized for distances inside the inner edge of the HZ. Note that the HZ limits assume Earth-sized planets with an initial water inventory equal to that of Earth for a 1-bar nitrogen atmosphere that is water-dominated at the inner edge and $CO_2$-dominated on the outer edge (following Kasting et al. 1993).

For Earth-like planets, the runaway greenhouse state (complete ocean evaporation) would be triggered when the planet's surface temperature reaches the critical temperature for water (647 K) (Kasting et al. 1993; Ramirez et al. 2014b), see section (3.3). For planets with smaller water inventories, the runaway greenhouse will be triggered at lower temperatures (see e.g. Kasting and Ackerman, 1986). We do not consider HZ limits for extremely dry (Abe et al. 2011, Zsom et al. 2014), or hydrogen-rich atmospheres (Pierrehumbert and Gaidos, 2011; see Kasting et al. 2013 for critical discussion of both limits) in this paper.

We calculate the effective stellar flux post-MS HZ limits (Fig.2; Table 1) and the corresponding post-MS HZ distances through the RGB and AGB (Fig.3; Table 2). Our 1D climate model calculates the limits for a grid of stars with effective temperatures between 2,600K and 10,000K (the model for 2,600K to 7,200K is discussed in detail in Kasting et al. 1993, Kopparapu et al. 2013,14, Ramses et al. 2014b). We use Bt-Settl modeled spectra (Allard 2003; 2007) to expand the temperature range of the climate model to 10,000 K for HZ calculations (following Kopparapu et al., 2013).

*2.3 Atmospheric Mass Loss Rates Model*

*2.3.1 Stellar Mass Loss Model*

We calculate stellar mass loss ($\dot{M}_{star}$ in $M_{sun}$/yr) using the Baud and Habing (1983) parameterization (1) for the AGB and a modified Reimers equation (2) for the RGB (Reimers, 1975, Vassiliadas and Wood 1993):

$$\dot{M}_{star\_RGB} = \frac{4}{3} x 10^{-13} \frac{L}{gR} \quad (1)$$

$$\dot{M}_{star\_AGB} = 4 x 10^{-13} \frac{M_i}{M} \frac{L}{gR} \quad (2)$$

Where $L$, $g$, and $R$, $M_i$, and $M$ are the stellar luminosity, gravity, radius, initial stellar mass at the start of the AGB, and stellar mass (in solar units), respectively. Resultant stellar masses, luminosities, and ages are given in Appendix A Table I. Stellar mass loss during the RGB phase reduces the remaining mass of the two smallest grid stars below 0.5 solar masses, too low to undergo the AGB phase (see e.g. Iben, 2013). Therefore we end our calculations at the end of the RGB, as predicted by the evolutionary models, or once a minimum mass of ~0.5 solar masses is reached (e.g., Sweigart and Gross, 1978; Iben, 2013), whichever comes first.

Studies have shown that Reimers' equation overestimates mass loss rates in the RGB, (Vassiliadas and Wood, 1993; Sweigert et al., 1990) leading Vasillidas and Wood (1993) to suggest multiplying Reimer's Law by 1/3 in order to obtain mass losses consistent with the HB morphology of RGB stars (Fusi Pecci and Renzini, 1976; 1978; Renzini et al., 1988), see equation (1). Although Reimer's equation was originally derived from a compilation of M-star data, G- and K-stars also exhibit similar mass loss rates (Reimers 1977, Vassiliadas and Wood 1993; Sweigert et al. 1990). Mass loss equations for A and F stars during the RGB are not available (e.g. Cranmer and Saar, 2011), therefore we assume here that those mass loss rates also follow Reimers' equation.

The Baud and Habing (1983) parameterization, see equation (2), includes the superwind near the end of the AGB, which predicts larger stellar mass losses than that those computed by Reimers' equation for this stellar phase in accordance with observations (ibid). The alternative Schroder and Cuntz (2005) parameterization gives similar answers to those given by the modified Reimers formula but is only valid for a relatively narrow $T_{eff}$ range (3,000 – 4,500 K; Schroder and Cuntz, 2005).

The Rosenfield et al. (2014) variant of Schroder and Cuntz's equation (2005) also yields stellar mass loss rates that are too strong for both the RGB and the early stages of the thermally pulsating AGB phase and is therefore not considered here. Recent models attempting to improve upon eqns. 1 and 2 (e.g. Cranmer and Saar, 2011) were not used because they require knowledge of unknown stellar parameters like surface X-ray fluxes (see discussion).

Table I in Appendix A shows that most of the mass loss for intermediate mass stars (~1 – 2 solar masses) occurs during the AGB (see also e.g. Villaver and Livio, 2007) and stars below solar mass would have too little mass (~0.5 Solar masses) to undergo the AGB phase (see i.e. Sweigart and Gross, 1978). Our K5 model reaches that mass at the end of the RGB whereas our added stellar mass loss causes our M1 model to achieve that mass slightly before the end of the RGB phase predicted by the Dartmouth models (Appendix A Table I). Note that most of the RGB mass loss for our M1 occurs in the last 1% of the RGB time frame (according to eqn. 1), therefore the exact time the RGB evolution of the M1 star ends does not influence the time a planet can spend within the HZ.

*2.3.2 Planetary Atmospheric Mass Loss Model*

The mixing layer formalism of Cantó and Raga (1991) shows that a flow with sound speed, $v_w$, and density, $p_w$, and a tangential velocity of the same order has an entrained mass flow $\dot{M} \approx \alpha p_w v_w$ (given entrainment efficiency, α). For a planetary atmosphere, we multiply the entrained mass flux by the surface of the leading hemisphere to estimate the planetary atmospheric mass loss per unit time, $M_a$, entrained by the stellar wind (following Zendejas et al. 2010 ):

$$\dot{M}_a \approx 2\pi R_p^2 \rho_w v_a; \quad (3)$$
$$v_a = \alpha v_w$$

Here, the atmospheric outflow velocity, $v_a$, is a function of the atmospheric entrainment efficiency, $\alpha$ (Cantó and Raga (1991)). The above expression can be combined with the relation $\dot{M}_{star} = 4\pi D^2 \rho_w v_w$ (where $D$ is the planetary orbital distance) to obtain equation (4):

$$\dot{M}_a = \left(\frac{R_p}{D}\right)^2 \frac{\dot{M}_{star} \alpha}{2} \quad (4)$$

Thus, the rate of planetary atmospheric mass loss is a function of both the orbital distance and the stellar mass loss rate. This expression can be combined with eqns. (1) and (2) to yield the planetary atmospheric mass loss as a function of stellar mass loss for the RGB phase eqn. (5) and AGB phase eqn. (6):

$$\dot{M}_{a\_RGB} = \frac{2}{3} x 10^{-13} \left(\frac{R_p}{D}\right)^2 \frac{L}{gR} \alpha \quad (5)$$

$$\dot{M}_{a\_AGB} = 2 x 10^{-13} \frac{M_i}{M} \left(\frac{R_p}{D}\right)^2 \frac{L}{gR} \alpha \quad (6)$$

By comparing laboratory experiments of plane, turbulent mixing layers, Cantó and Raga (1991) determined that α = 0.03. Bauer and Lammer (2004) suggested an upper value of α = 0.3 for Venus. Planets with Earth-like atmospheres should be less turbulent than Venus, thus we use an intermediate value for α of 0.2 for our calculations and explore the sensitivity of the results to this parameter (see discussion and Appendix B). We terminate our calculations either when planetary surface pressures become smaller than 0.25 bar, or at the end of the AGB, whichever comes first.

*2.3.3 Planetary Orbital Radius Expansion Model*

We assume that planets are far enough away from the expanding star that orbital dissipation arising from tidal forces are negligible. As long as the planet's orbital distance is larger than the stellar radius of the expanding star, and tidal disruption effects are negligible, the orbital variation can be approximated by eqn.(7) (e.g. Zahn, 1977, Villaver and Livio, 2007):

$$\frac{1}{D}\frac{dD}{dt} = -\frac{1}{M_*}\frac{dM_*}{dt} \qquad (7)$$

Here, $M_*$ is the mass of the star at the end of the main-sequence. This expression is integrated to give:

$$D(t) = D_o \frac{M_*}{M_*(t)} \qquad (8)$$

Where $D_o$ is the initial orbital distance and $M_*(t)$ is the time-dependent stellar mass loss during the RGB and AGB and $D(t)$ is the orbital distance of the planet at time $t$.

*2.3.4 Super-Earths and Super-Moon Models*

We model 0.5 $M_{Earth}$ super-moons as well as 5 and 10 $M_{Earth}$ super-Earths for comparison to Earth-analog planets. We derive the planetary radius using the parameterization given in Valencia et al. (2007) for a planet with a bulk $H_2O$ content of 0% (which most closely approximates the composition of the Earth), yielding 0.83, 1.61, and 1.83 Earth radii, respectively, in order of increasing mass. The resulting atmospheric masses are scaled by the volume of the planets, producing planetary atmospheric masses ~ 0.57, 3.51, and 6.12 times that of the Earth, respectively, in order of increasing planetary mass.

**3 RESULTS**

*3.1 The Post-Main-Sequence Habitable Zone Limits*

The HZ limits evolve through the stellar post-MS both in effective stellar flux on top of the planet's atmosphere, $S_{eff}$, (Fig. 2) as well as orbital distance from the star (Fig. 3) due to the star's changing SED and luminosity, respectively. We derive a parameterization (eq. 9) to compute post-main-sequence HZ distances for grid stars old enough to be currently on the post-main-sequence (Sun – A5) using the constants given in Appendix A Table ID. Our parameterization only includes the slowly changing portion of the RGB, during which most planets retain their atmospheres.

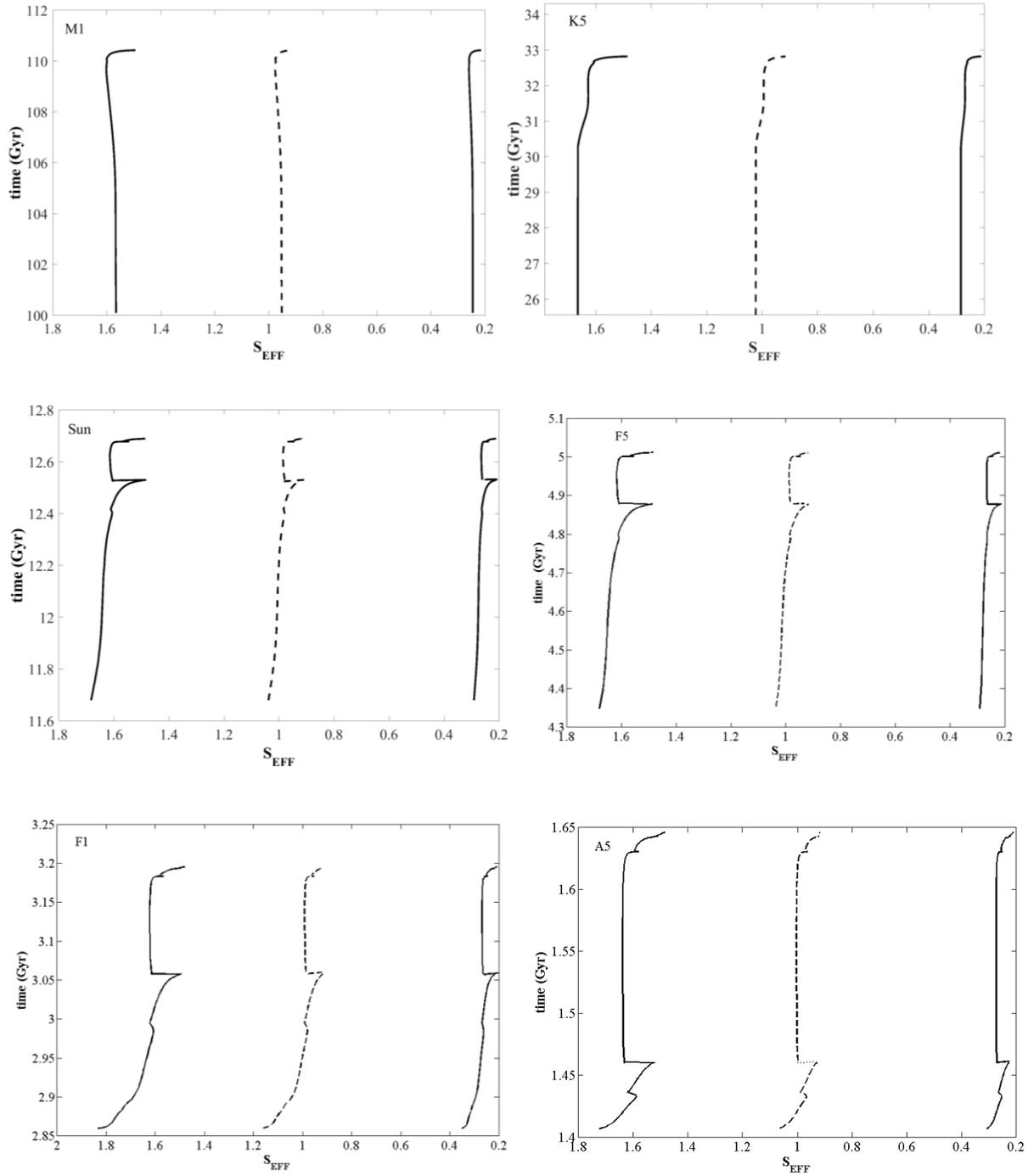

*Figure 2*: Post-MS HZ boundaries for the empirical HZ (solid) and modeled 3D inner boundary (dashed) for A5 – M1 stars shown in effective stellar flux $S_{eff}$ (stellar flux on top of Earth's atmosphere).

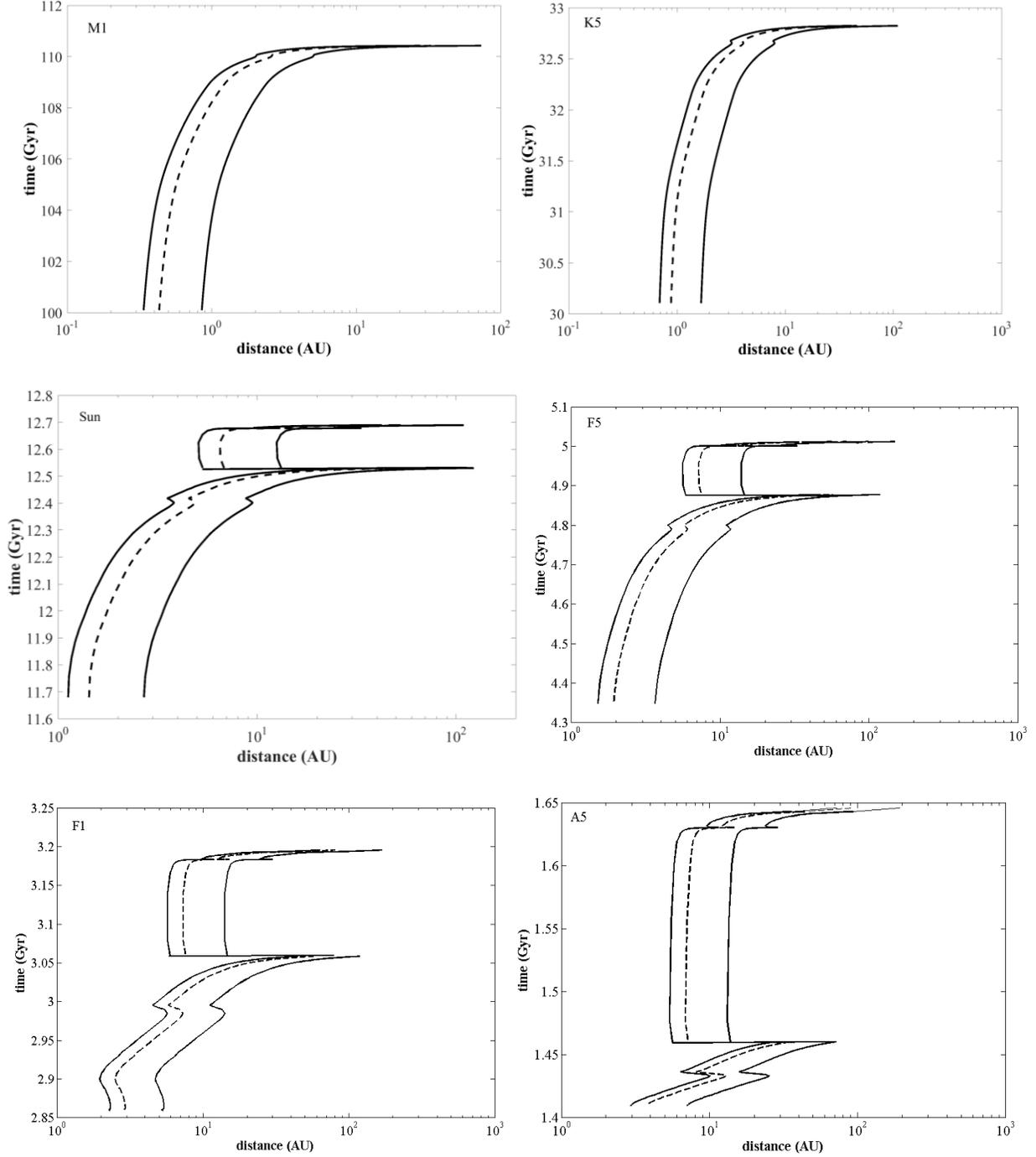

***Figure 3****: Post-MS-HZ boundaries for the empirical HZ (solid), and modeled 3D inner edge (dashed) for A5 – M1 stars shown in orbital distance.*

$$D = at^4 + bt^3 + ct^2 + dt + e \qquad (9)$$

Where D is the distance (in AU) and *t* is stellar age for solar metallicity stars (in Gyr).

The effective stellar flux at the boundaries of the post-MS-HZ as shown in Fig. 2 decreases during the RGB by a maximum of ~22% for the F1 and 5% for the M1 stellar type. These two stellar types bracket the maximum and minimum luminosity change for our grid of stars, respectively. The corresponding increases of $S_{eff}$ during the Helium-flash and decreases during the following AGB, respectively, are 42% and 69% for the M1 and 8% and 8% for the F1 stellar type. Note that the stellar evolution timescales differ for the different stars (Figures 2 – 3).

The orbital distance of the post-MS-HZ limits changes through time (Fig.3; Table 2). The inner and outer edges of the post-MS HZ for an A5 star is initially located at 3 and 7 AU respectively, at the beginning of the RGB before moving to 28 and 72 AU respectively by the end of the RGB (total duration: ~ 50 Myr). After the helium flash, the inner and outer edges are located at 6 and 14 AU respectively at the start of the AGB, extending outward to 72 and 192 AU by the end of the AGB (total duration: ~ 183Myr).

Our Sun's inner and outer edges of the post-MS HZ are initially located at 1.3 and 3.3 AU before expanding outward to 46 and 123 AU by the end of the RGB (total duration: ~ 850 Myr). During the AGB the post-MS HZ edges move outward from 5 and 13 AU to 39 and 110 AU respectively (total duration: ~ 160 Myr).

For the coolest grid star (M1), the post-MS HZ edges are initially located at 0.3 and 0.9 AU and increase out to 13 and 34 AU by the end of the RGB (total duration: ~ 9 Gyr).

**Table 1: Post-main-sequence habitable zone distance limits in effective stellar flux ($S_{eff}$) on top of the planet's atmosphere**

| Star | Beginning RGB (inner) | Beginning RGB (outer) | End RGB (inner) | End RGB (outer) | Beginning AGB (inner) | Beginning AGB (outer) | End AGB (inner) | End AGB (outer) |
|---|---|---|---|---|---|---|---|---|
| M1* | 1.549 | 0.245 | 1.493 | 0.223 | - | - | - | - |
| K5* | 1.676 | 0.286 | 1.485 | 0.21 | - | - | - | - |
| Sun | 1.683 | 0.29 | 1.482 | 0.209 | 1.61 | 0.261 | 1.487 | 0.211 |
| F5 | 1.681 | 0.290 | 1.488 | 0.212 | 1.61 | 0.261 | 1.499 | 0.217 |
| F1 | 1.832 | 0.35 | 1.492 | 0.214 | 1.62 | 0.264 | 1.482 | 0.218 |
| A5 | 1.723 | 0.307 | 1.53 | 0.228 | 1.63 | 0.271 | 1.485 | 0.211 |

*M1 and K5 stars do not reach the AGB (see text).

**Table 2: Post-main-sequence habitable zone distance limits in AU**

| Star | Beginning RGB (inner) | Beginning RGB (outer) | End RGB (inner) | End RGB (outer) | Beginning AGB (inner) | Beginning AGB (outer) | End AGB (inner) | End AGB (outer) |
|---|---|---|---|---|---|---|---|---|
| M1* | 0.34 | 0.86 | 13 | 33.5 | - | - | - | - |
| K5* | 0.69 | 1.67 | 40.24 | 108.10 | - | - | - | - |
| Sun | 1.3 | 3.3 | 46.18 | 123 | 5.34 | 13.24 | 39.1 | 109.5 |
| F5 | 1.52 | 3.66 | 45.9 | 121.6 | 5.88 | 14.55 | 55.63 | 147.7 |
| F1 | 2.27 | 5.21 | 44.6 | 117.6 | 5.96 | 14.73 | 62.95 | 167.7 |
| A5 | 2.95 | 7.07 | 27.4 | 71.7 | 5.64 | 13.8 | 72.24 | 191.9 |

*M1 and K5 stars do not reach the AGB (see text)

The maximum amount of time that a planet can remain in the post-MS HZ varies with stellar spectral class (Table 3) (as well as metallicity), see section 3.2. A planet orbiting the coolest grid star (M1) can remain in the post-MS HZ for about 9 Gyr, assuming solar metallicity (Table 3). In contrast, a planet orbiting the hottest grid star (A5) can only remain in the HZ for up to 40 million years. A planet orbiting a post-MS Sun can reside in the HZ for ~ 500 million years, which is comparable to the amount of time for life to evolve on the Earth (~700 Myr; Mojzsis & Arrhenius, 1997). Note that the above quoted numbers are conservative estimates because they assume that the planet's orbital radius stays constant with time. However, these times increase as the star loses mass and the planets' orbits therefore move outwards (see section 3.3 and Appendix A Tables I - II). Also stellar metallicity can extend that time significantly.

*3.2 Effects of Metallicity on HZ boundaries*

Stars with higher metallicity evolve more slowly and therefore the time a planet at a certain distance remains in the habitable zone is longer than for a star with lower metallicity (see Danchi & Lopez 2013). Higher metallicity stars have proportionately less H to fuse, which slows down nuclear burning.

We show the influence of metallicity on the location as well as evolution of the post-MS-HZ boundaries by comparing the same star type with low (Fe/H = -0.5) and high (Fe/H = 0.5) metallicity for all grid stars (Fig. 4). These metallicity values are those assumed at a stellar age of zero, which subsequently evolves as the star ages. We find that the time in the post-MS-HZ is about half as long for a planet orbiting the low metallicity star as it is for one orbiting the high metallicity star (see Figure 4). Although stellar metallicity changes the parameters of the evolving star, both $S_{eff}$ as well as the orbital distance at the limits of the post-MS HZ boundaries change. For stars with high metallicity, the post-MS orbital distance limits are smaller compared to host stars with lower metallicity (see Fig. 4). Thus, for stars with non-solar metallicities, the parameterization for the distance of the post-MS HZ, eqn. (9), as a function of the star's age has to be adapted according to their slower evolution.

*3.3 Post-main-sequence atmospheric erosion of Earth-mass planets at the Mars-, Jupiter-, Saturn-, and Kuiper Belt locations*

Previous studies have argued that life may also initiate during the relatively short period a planet spends in the post-MS HZ (e.g. Lopez et al., 2005; Danchi and Lopez, 2013). However, the HZ is not the distance where life can evolve but is used to remotely detect signs of life if it exists. Therefore life may have also initiated before the stellar post-MS on planets outside the MS HZ, thriving subsurface (e.g. underneath an ice layer). Once the post-MS HZ moves outward, such worlds would get heated, potentially uncovering hidden ecosystems, which could result in remotely detectable atmospheric biosignatures. Our Sun's post-MS-HZ moves outward (except between the RGB and AGB) beyond the Kuiper belt (see also Stern 1990) to the outer regions of the solar system.

Melting would occur on icy outer bodies (i.e., Europa, Enceladus, Ganymede, Pluto). That could possibly trigger a water-based hydrologic cycle on these potentially habitable worlds.

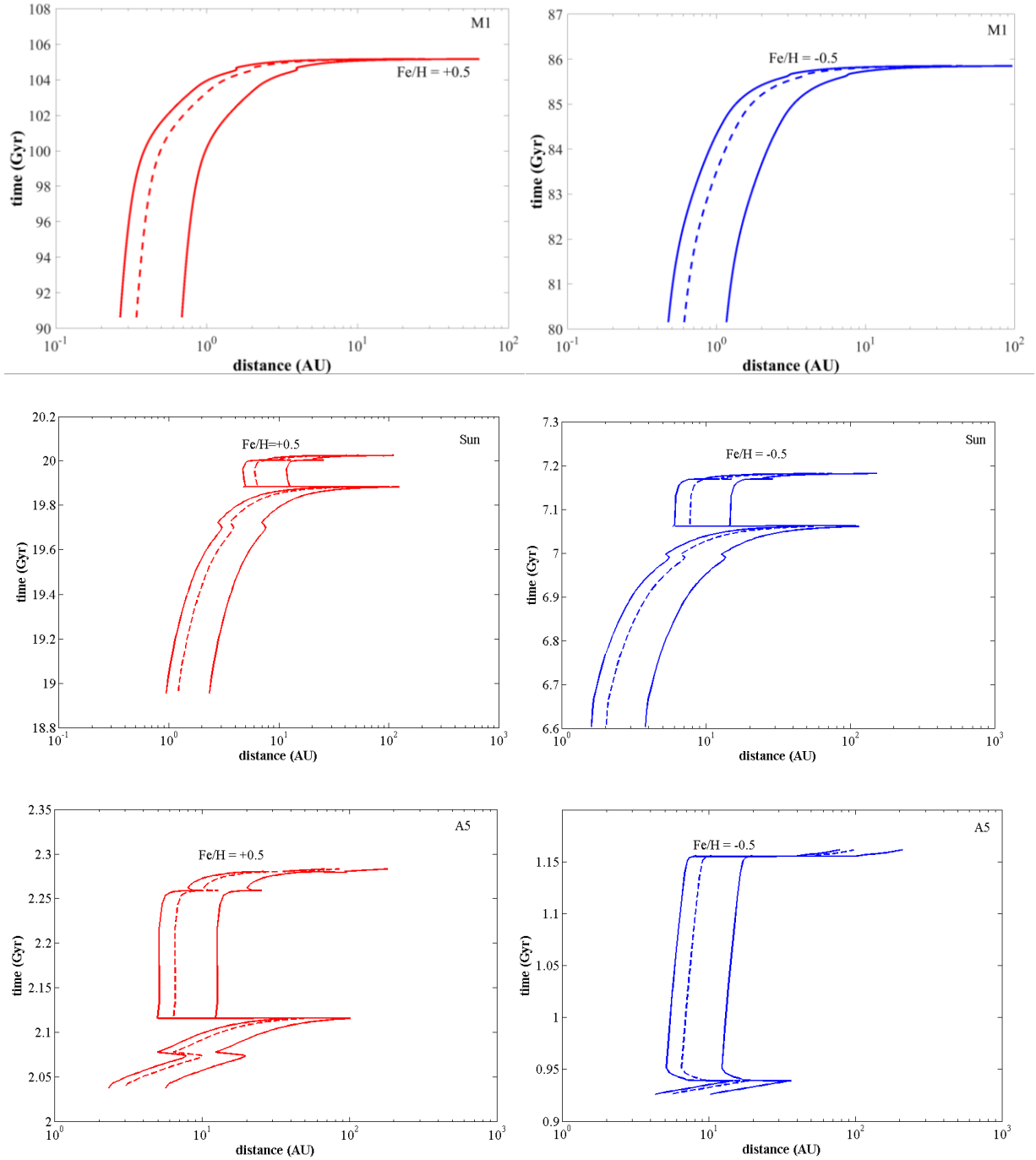

*Figure 4: Post-MS HZ boundaries for high stellar metallicity (Fe/H= 0.5; left panel) and lower stellar metallicity (Fe/H = -0.5; right panel) for the Sun (middle) as well as the coolest (M1, top) and hottest (A5, bottom) grid stars. M1 boundaries are calculated the star reaches a minimum mass of 0.5 solar masses, at the tip of the RGB.*

The icy moons in our own solar system are not massive enough to maintain a dense atmosphere if heated. Therefore they should also not support an effective temperature transport between the planetary day and night side (e.g., Joshi et al., 1997; Joshi, 2003; Leconte et al. 2013b).

But in other planetary systems, more massive moons or even Earth-mass planets could be located at equivalent distances. Therefore we use the location of Mars, Jupiter, Saturn and the Kuiper Belt in our own Solar System to calculate how long Earth-mass bodies could retain their atmospheres for all grid stars (Figures 5 - 6). We scale the initial atmospheric surface pressure of the planet with gravity (following Kaltenegger et al. 2013). We include the effect that planetary orbital distances increase as post-MS stars lose their stellar mass, including the erosion of the planet's atmosphere due to these resulting stellar winds (Fig. 5 - 6, Appendix A Tables III - IV)(e.g. Lorenz et al., 1997).

We terminate our calculations (indicated by a filled blue sphere in Fig. 5 and 6) when planetary atmospheres losses lead to atmospheric surface pressure of 0.25 bar or less. The orbital expansion for Earth-mass planets orbiting the grid host stars during the RGB and AGB stages are also shown in Fig. 5-6, respectively. Our calculations initially put the planet at Mars-, Jupiter-, Saturn- , and Kuiper belt-equivalent orbital distances (scaled to the stellar fluxes at 1.52, 5.2, 9.5, and 30 AU in our solar system).

More planetary atmospheres within the post-MS HZ of high mass stars survive through the stellar RGB and AGB phase than those within the post-MS HZ of lower mass stars because of the higher mass loss during the post-MS phase for the latter. Except for the stellar spectral type A5, all planets at the Mars-equivalent distance lose their atmospheres during the RGB phase. None of the planetary atmospheres at the Mars-equivalent distance are able to survive through the end of the AGB. However, planetary atmospheres located at least as far as Jupiter's equivalent distance from orbiting stars at least as massive as an F5 stellar type are able to retain their atmospheres through the end of the AGB.

Cool stars have longer post-MS phases and therefore planets around our coolest grid star, M1, spend the longest time in the post-MS HZ, 9 Gyr for solar metallicity. Planets at Jupiter- and Saturn-equivalent orbital distances remaining in the post-MS HZ for up to 5.8 Gyr and 2.1 Gyr, respectively (Table 3). For our Sun, a planet at Jupiter's distance could remain in the post-MS HZ for up to 370 Myr during the stellar post-MS, whereas equivalent planets at Mars' and Saturn's distances stay in the post-MS HZ for ~200 Myr. Planets orbiting an A5 star only stay in the post-MS HZ for tens of millions of years.

In all cases, objects at the Kuiper belt-location are also eventually heated but this heating is rather short-lived (200 and 100 million years for planets around an M1 and an A5 spectral type respectively) as it coincides with the end of the RGB.

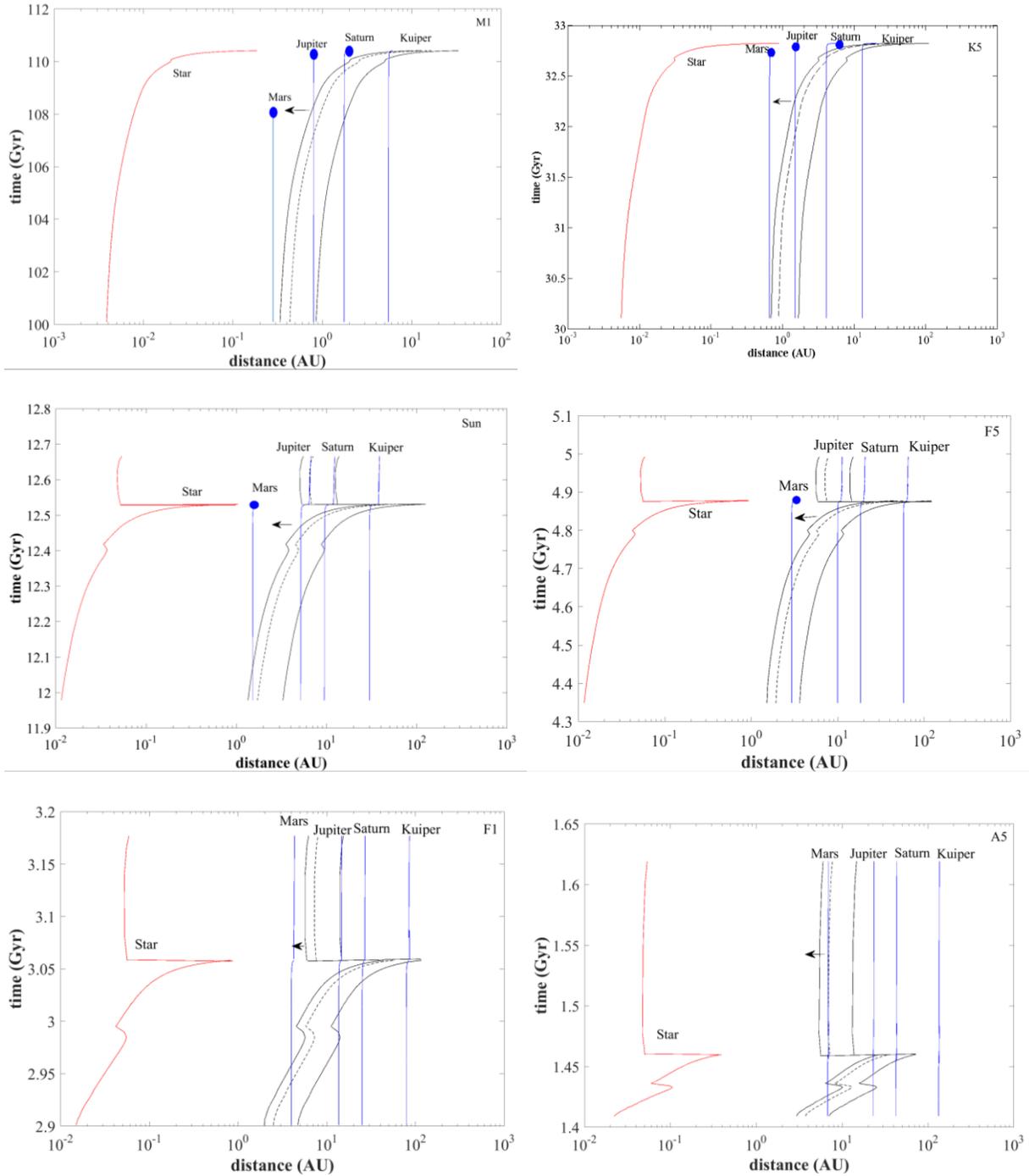

*Figure 5: Temporal evolution of stellar radius (left) and RGB + HB HZ (black solid and dashed lines) as well as equivalent distance of Mars-, Jupiter-, Saturn-, and Kuiper belt Earth-mass planets in our Solar System (blue straight lines) for all grid stars. Blue dots indicate planetary surface pressures smaller than 0.25 bar. The runaway greenhouse (left arrow) is triggered inside the inner edge of the HZ.*

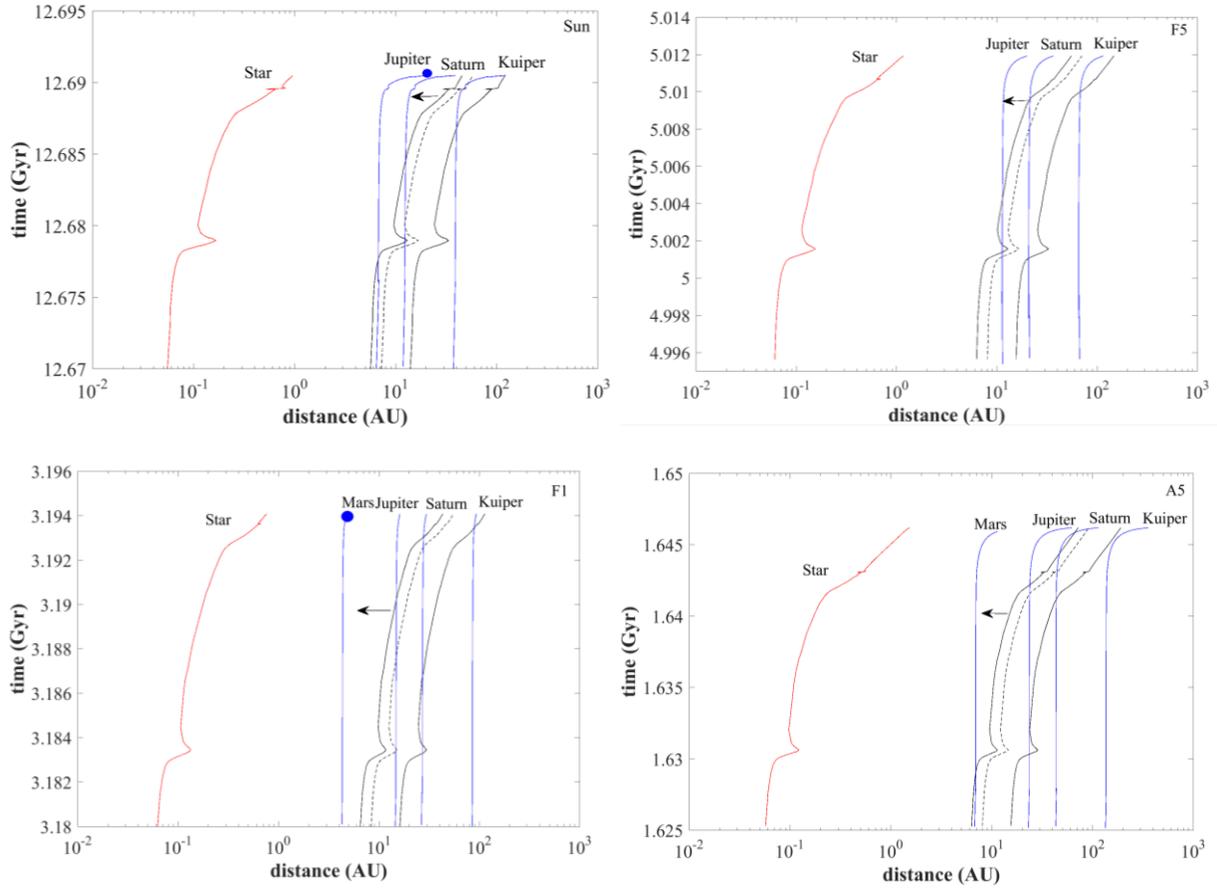

*Figure 6:* *Temporal evolution of stellar radius (left) and AGB HZ (black solid and dashed lines) as well as equivalent distance of Mars-, Jupiter-, Saturn-, and Kuiper belt Earth-mass planets in our Solar System (blue straight lines) for the Sun, F5, F1, and A5 grid stars. Blue dots indicate planetary surface pressures smaller than 0.25 bar. The runaway greenhouse (left arrow) is triggered inside the inner edge of the HZ.*

**Table 3: Time (in Gyr) a planet spends in the post-MS HZ***

| Star | Max time in HZ | Mars | Jupiter | Saturn | Kuiper Belt |
|------|----------------|------|---------|--------|-------------|
| M1   | 9.00           | -    | 5.80    | 2.1    | 0.50        |
| K5   | 2.10           | -    | 2.10    | 0.45   | 0.06        |
| Sun  | 0.5            | 0.1  | 0.37    | 0.21   | *           |
| F5   | 0.39           | 0.35 | 0.23    | 0.03   | *           |
| F1   | 0.07           | 0.07 | 0.06    | 0.02   | *           |
| A5   | 0.21           | 0.20 | 0.035   | 0.01   | *           |

\* *Stars* indicate that those planets do not spend more than 1 Myr in the post-MS HZ,
- *Dashes* indicate that these planets are not located in the post-MS HZ during the RGB and AGB phase of the star.

## 3.4 Post-Main-Sequence Erosion of Super-Moon to Super-Earth Atmospheres

Higher stellar mass loss rates for low mass stars increase atmospheric mass loss for their planets. Although the increase in stellar mass loss would also result in more pronounced orbital radius expansion for these planets (Appendix A Table II), the maximum times a planet can spend in the post-MS HZ given in Table 3 are negligibly affected because 1) the bulk of the radial expansion occurs near the tail end of the AGB, and 2) this AGB expansion occurs once the planet is outside the HZ in most cases (Figure 6). The orbital distances to receive the same stellar flux as planets in our own Solar System are correspondingly farther away for more massive stars, reducing atmospheric mass loss for their planets.

Tables III and Table IV in Appendix A show the atmospheric mass loss for planets with 0.5, 1, 5, and 10 Earth masses for the RGB and the AGB phase of their host grid star respectively.

Super-moons are only slightly more susceptible to atmospheric loss than Earth-mass planets and super-Earths. All modeled planets at the Mars-equivalent distance lose their atmospheres at the end of the RGB for all grid stars of solar mass or smaller (Appendix A Table III). For the smallest grid star (0.5 initial solar mass), planets out to Saturn's orbit lose their atmosphere during the RGB. During the stellar AGB phase, all planets at the Mars-equivalent distance lose their atmospheres for grid stars with 1 - 1.5 initial solar masses (Appendix A Table IV).

More massive planets also retain their atmospheres for longer periods of time than do less massive ones (see Fig. 5 - 6). Super-Earths are resistant to atmospheric loss compared to Earth-mass planets. At larger distances, super-moon atmospheres located at the Kuiper-belt equivalent distance survive through the entire AGB phase for all grid stars.

## 3.5 Directly Imaged planets

We use known exoplanets to compare their orbital distance to the post-MS HZ to estimate detection capabilities for such planets. The planets currently detected at large distances are young, massive worlds that have not yet entered the post-MS HZ stage. But they show that the post-MS HZ is populated by known young planets. They also showcase the capabilities of current detection methods. We use three known exoplanet systems with detected planets located at orbital distances between 9 AU to 110 AU and plotted their orbits on top of the post-MS HZ (see Fig. 7). The planets are shown as pointed lines at their current orbital distance in the plot. Those orbital distance lines intersect with the post-MS HZ distance for all three systems. Note, that because the exact characteristics of these planets are unknown, we could not estimate the evolution of their mass and orbital distance through the star's evolution into the post-MS.

Beta pictoris (spectral class: A6) is ~10 – 20 million years old and has a mean surface temperature of ~ 8,050 K and luminosity 8.7 times that of the Sun (Crifo et al., 1997; Zuckerman et al., 2001; Gray et al., 2006). Beta pictoris b has a mass between 4 and 11 Jupiter masses and a radius 65% larger than Jupiter's (Currie et al., 2013), orbiting at a distance of ~ 9 AU from Beta pictoris. Fig. 7 (top) shows the post-MS HZ orbital distance for the host star.

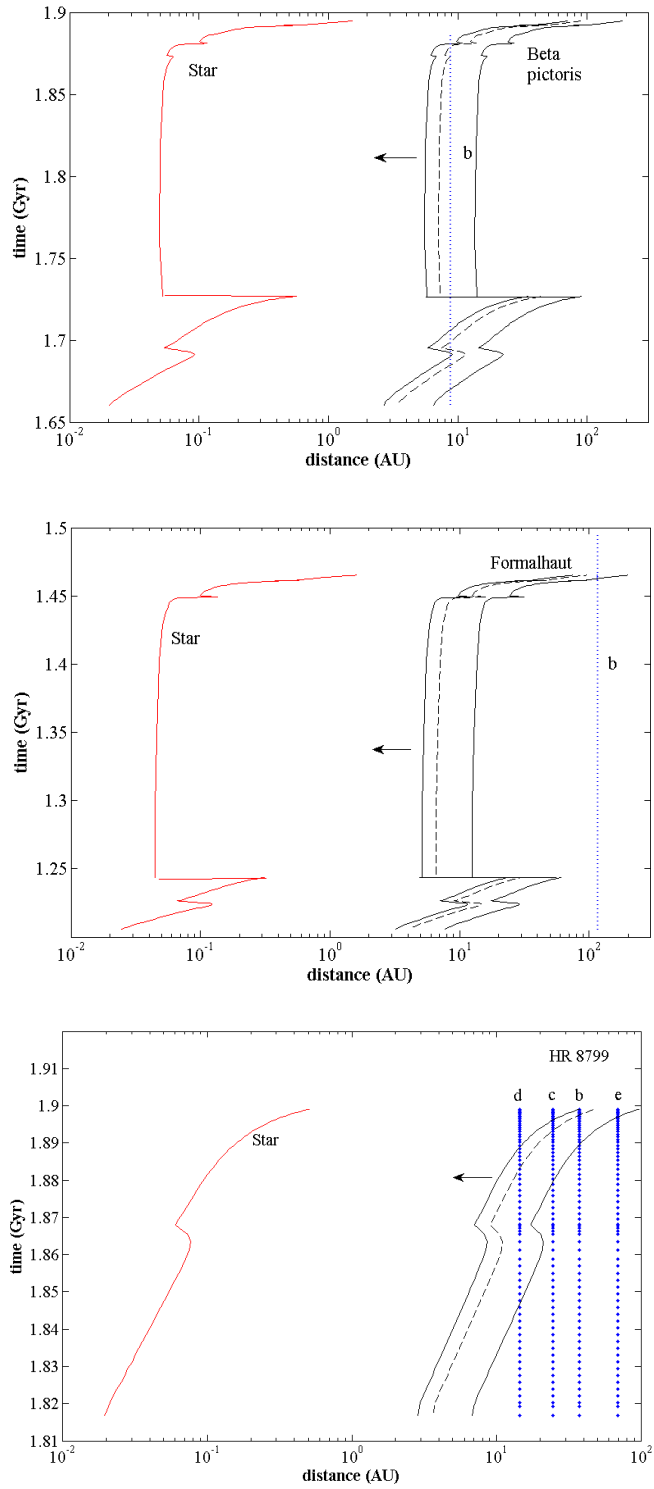

*Figure 7: Comparison of known young directly imaged exoplanets (blue dots) with the radius (left red line) and post-MS HZ distance of their host stars (black lines). The runaway greenhouse (arrow) is triggered inside the inner edge of the HZ.*

Fomalhaut (spectral class: A3) is ~440 million years old and has a mean surface temperature of ~ 8,590 K and luminosity 16.63 times that of the Sun (Mamajek, 2012). Fomalhaut b orbits at a distance ~ 110 AU from its parent star and has a mass estimated between 0.075 and 3 Jupiter masses (Kalas et al., 2013). Fig. 7 (middle) shows the post-MS HZ for the host star.

HR 8799 (spectral class: A5 V; λ Boo} is ~ 30 million years old and has a mean surface temperature of 7,430 K and luminosity 4.92 times that of the Sun (Gray and Kaye, 1999). The planets e, d, c, and b orbit ~ 14.5, 24, 38, and 68 AU from their parent star, with d, c, and b measured to have a planetary radius of 1.2 times that of Jupiter (Marois et al., 2010). The planets e, d, and b each have masses equal to 7 times that of Jupiter, with b having a mass 5 times that of Jupiter (Marois et al., 2010). Fig. 7 (bottom) shows the post-MS HZ for the host star.

**DISCUSSION**

Our results also explore planetary atmospheric mass loss trends as a function of stellar mass loss for a range of planetary masses. Our parameterizations make no distinction between the pre-dust-driven and dust-driven mass loss stages of the AGB. This is because a dust-driven mass model requires specific knowledge of observational parameters, including galactic long period variables and individual stellar compositions that are beyond the scope of the general analysis here. Moreover, sticking coefficients for certain minerals (e.g. pyroxene) are unknown and dust extinction properties are also poorly understood, leading to major uncertainties in stellar mass loss rates for dust driven models (Ferrarotti and Gail, 2002). In addition to these uncertainties, accurate and precise determination of stellar parameters continue to be a fundamental problem in stellar models. A recent study by Valle et al. (2014) analyzed the impact on mass and radius determination of the uncertainty in the input physics, mixing-length value, initial helium abundance, and microscopic diffusion efficiency adopted in stellar model computations. They find that errors and biases with the estimation grids for these models yield greater errors than can be due to statistical differences. The results here can be updated once better estimates for stellar mass loss in the post-MS phase of the star become available.

The entrainment efficiency, $\alpha$, is set here to 0.2 to estimate the planetary atmospheric mass loss. The range estimated in the literature spans from 0.03 to 0.3 (see the sensitivity study to this parameter in Appendix B). If the entrainment efficiency is very low (0.03), the planetary atmospheric mass loss would be about a factor of 10 lower than in our calculations and planets would keep their atmospheres longer even at small orbital distances. If the entrainment efficiency, is higher, 0.3 instead of 0.2, planets would lose their atmospheres faster than calculated here. Our sensitivity study shows that even at $\alpha = 0.3$, Earth-mass planets located at Saturn's distance in our solar system retain at least part of their atmospheres at the end of the AGB (Appendix B).

None of the cool late K and M stars have yet reached the post-MS phase, making the lifetime in the post-MS HZ for cool stars a prediction, not an observable quantity. Nevertheless with M stars being the most abundant stars, they are part of the exploration of parameter space to expand our understanding of the detectability of habitable planets and maintenance of liquid water on the surface of a planet.

## CONCLUSION

Well over 99.9% of all the water in our solar system lies in the region beyond the zero-age main-sequence ice line (Stern, 2003), making this region an interesting prospect for potential biological evolution. Such life could become detectable in the atmospheres of such planets during post-MS stellar evolution, in the post-MS HZ.

We calculate the post-MS HZ for a grid of stars from 3,700K to 10,000K (~M1 to A5 stellar types) for different stellar metallicities. We derive a parameterization (eqn. 9) to compute post-main-sequence HZ distances for grid stars old enough to be currently on the post-main-sequence (Sun – A5). Planetary atmospheric erosion during the post-main-sequence is mainly due to high stellar winds produced by the stellar mass loss, which can erode planetary atmospheres. Super-Moons to super-Earths' atmospheres can survive the RGB and AGB phase of their host star – except for planets on close-in orbits. Even super-moons survive at a Kuiper-belt equivalent distance for all grid stars to the end of the AGB phase.

Planets can stay in the post-MS HZ between 200 million and 9 Gyr for the hottest and coolest grid stars respectively. That time increases further with increasing stellar metallicity. Although post-MS HZ lifetimes for massive stars are relatively short, life that was previously undetectable during the MS or pre-MS can become remotely detectable as stellar luminosities rise.

Known directly imaged exoplanets orbit their stars at comparable distances to the post-MS HZ. The planets currently detected at large distances are young and hot planets, and therefore the post-MS HZ, does not apply to these Systems. But they show that the post-MS HZ distance is populated by known young planets and that potentially rocky planets at an older stage could be detected in the post-MS HZ in the near future.

**Acknowledgements:** The authors thank the anonymous referee for constructive comments and helpful insights. The authors acknowledge support by the Simons Foundation (SCOL # 290357, Kaltenegger) and the Carl Sagan Institute.

# APPENDIX

## APPENDIX A

Stellar and planetary orbital evolution calculations are tabulated in Tables I and II, respectively. Atmospheric loss estimates for the RGB and AGB are given in Tables III and IV, respectively.

**TABLE I.** * Note that the K5 and M1 stars do not reach the AGB (see text).

A) Shows the stellar mass (in solar masses) at the start of the RGB, end of the RGB, and end of the AGB phases

| Star | M1 | K5 | Sun | F5 | F1 | A5 |
|---|---|---|---|---|---|---|
| **Stellar mass (start RGB)** | 0.50 | 0.75 | 1.00 | 1.30 | 1.50 | 1.90 |
| **Stellar mass (end RGB)** | 0.45 | 0.48 | 0.80 | 1.17 | 1.41 | 1.88 |
| **Stellar mass (end AGB)** | * | * | 0.5 | 0.65 | 0.64 | 0.7 |

B) Shows the stellar luminosity (in solar luminosities) at the start of the RGB and AGB as well as the end of the RGB and AGB phases, respectively.

| Star | M1 | K5 | Sun | F5 | F1 | A5 |
|---|---|---|---|---|---|---|
| **Stellar luminosity (start RGB)** | 0.18 | 0.80 | 2.13 | 3.88 | 6.57 | 14.72 |
| **Stellar luminosity (end RGB)** | 256 | 2535 | 3162 | 2988 | 2974 | 1379 |
| **Stellar luminosity (start AGB)** | - | - | 50 | 61 | 69 | 65 |
| **Stellar luminosity (end AGB)** | - | - | 2556 | 4597 | 5871 | 7750 |

C) Shows the stellar ages (in Gyr) at the start of the RGB and AGB as well as the end of the RGB and AGB phases, respectively.

| Star | M1 | K5 | Sun | F5 | F1 | A5 |
|---|---|---|---|---|---|---|
| **Stellar age (start RGB)** | 100.1 | 30.1 | 11.98 | 4.35 | 2.9 | 1.41 |
| **Stellar age (end RGB)** | 110.4 | 32.8 | 12.53 | 4.88 | 3.06 | 1.46 |
| **Stellar age (start AGB)** | - | - | 12.67 | 5 | 3.18 | 1.63 |
| **Stellar age (end AGB)** | - | - | 12.69 | 5.01 | 3.2 | 1.65 |

D) Shows the constants to compute the empirical post-MS HZ boundaries using eqn. (9). The inner edge is the Recent Venus column, the outer edge, the Early Mars column. An alternative limit for the inner edge from 3D models is also shown. Note that the outer limits agree in the 3D and 1D model and are therefore not given in separate columns. The corresponding post-MS ages when the equation is valid are during the RGB (except for the last ~2% of RGB time).

| Sun **Constant** | **Recent Venus** | **3D inner limit** | **Early Mars** |
|---|---|---|---|
| A | 132 | 169.8 | 337.6 |
| B | -6385 | -8213 | $-1.633 \times 10^4$ |
| C | $1.158 \times 10^5$ | $1.49 \times 10^5$ | $2.963 \times 10^5$ |
| D | $-9.341 \times 10^5$ | $-1.201 \times 10^6$ | $-2.389 \times 10^6$ |
| E | $2.824 \times 10^6$ | $3.633 \times 10^6$ | $7.226 \times 10^6$ |

| F5 Constant | Recent Venus | 3D inner limit | Early Mars |
|---|---|---|---|
| A | 255.6 | 327.9 | 642.8 |
| B | -4605 | -5908 | $-1.158 \times 10^4$ |
| C | $3.111 \times 10^4$ | $3.992 \times 10^4$ | $7.822 \times 10^4$ |
| D | $-9.3444 \times 10^4$ | $-1.199 \times 10^5$ | $-2.349 \times 10^5$ |
| E | $1.052 \times 10^5$ | $1.35 \times 10^5$ | $2.644 \times 10^5$ |

| F1 Constant | Recent Venus | 3D inner limit | Early Mars |
|---|---|---|---|
| A | $2.374 \times 10^4$ | $2.924 \times 10^4$ | $5.234 \times 10^4$ |
| B | $-2.798 \times 10^5$ | $-3.446 \times 10^5$ | $-6.164 \times 10^5$ |
| C | $1.237 \times 10^6$ | $1.524 \times 10^6$ | $2.723 \times 10^6$ |
| D | $-2.433 \times 10^6$ | $-2.996 \times 10^6$ | $-5.35 \times 10^6$ |
| E | $1.794 \times 10^6$ | $2.209 \times 10^6$ | $3.942 \times 10^6$ |

| A5 Constant | Recent Venus | 3D inner limit | Early Mars |
|---|---|---|---|
| A | $2.167 \times 10^7$ | $3.319 \times 10^7$ | $6.407 \times 10^7$ |
| B | $-1.485 \times 10^8$ | $-1.883 \times 10^8$ | $-3.635 \times 10^8$ |
| C | $3.16 \times 10^8$ | $4.008 \times 10^8$ | $7.733 \times 10^8$ |
| D | $-2.989 \times 10^8$ | $-3.79 \times 10^8$ | $-7.312 \times 10^8$ |
| E | $1.06 \times 10^8$ | $1.344 \times 10^8$ | $2.593 \times 10^8$ |

**TABLE II**

A) Orbital evolution of planets at Mars equivalent distances in our Solar System

| Stellar mass | 0.5 | 0.75 | 1 | 1.3 | 1.5 | 1.9 |
|---|---|---|---|---|---|---|
| Orbital Radius initial | 0.281 | 0.658 | 1.52 | 2.94 | 4 | 6.79 |
| Orbit Radius end RGB | 0.312 | 1.02 | 1.89 | 3.26 | 4.27 | 6.88 |
| Orbit Radius end AGB | - | - | 5.58 | 5.87 | 11.28 | 18.38 |

B) Orbital evolution of planets at Jupiter equivalent distances in our Solar System

| Stellar mass | 0.5 | 0.75 | 1 | 1.3 | 1.5 | 1.9 |
|---|---|---|---|---|---|---|
| Orbital Radius initial | 0.957 | 2.24 | 5.2 | 10 | 13.68 | 23.13 |
| Orbit Radius end RGB | 0.889 | 3.48 | 6.48 | 11.1 | 14.6 | 23.43 |
| Orbit Radius end AGB | - | - | 19.09 | 19.98 | 32.16 | 62.62 |

C) Orbital evolution of planets at Saturn equivalent distances in our Solar System

| Stellar mass | 0.5 | 0.75 | 1 | 1.3 | 1.5 | 1.9 |
|---|---|---|---|---|---|---|
| Orbital Radius initial | 1.76 | 4.11 | 9.53 | 18.38 | 25.1 | 42.46 |
| Orbit Radius end RGB | 1.956 | 6.39 | 11.86 | 20.41 | 26.79 | 43.01 |
| Orbit Radius end AGB | - | - | 35 | 36.7 | 59 | 114.95 |

D) Orbital evolution of planets at Kuiper Belt equivalent distances in our Solar System

| Stellar mass | 0.5 | 0.75 | 1 | 1.3 | 1.5 | 1.9 |
|---|---|---|---|---|---|---|
| Orbital Radius initial | 5.53 | 12.95 | 30 | 57.9 | 79 | 133.56 |
| Orbit Radius end RGB | 6.146 | 20.14 | 37.34 | 64.29 | 84.318 | 135.292 |
| Orbit Radius end AGB | - | - | 110.2 | 115.7 | 185.7 | 361.59 |

**TABLE III**

Atmospheric mass loss during the RGB for all grid stars (* denotes atmospheric surface pressure below 0.25 bar, where we terminate the calculations). All values are given in % of remaining atmosphere.

A) Atmospheric loss for super-moons with 0.5 Earth mass planets with a 0.57 bar initial surface pressure

| Star RGB | Mars | Jupiter | Saturn | Kuiper Belt |
|---|---|---|---|---|
| 0.5 | * | * | * | 87.7 |
| 0.75 | * | * | * | 91.2 |
| 1 | * | 50.9 | 86.0 | 98.2 |
| 1.3 | * | 89.5 | 96.5 | 100.0 |
| 1.5 | 54.4 | 96.5 | 98.2 | 100.0 |
| 1.9 | 96.5 | 100.0 | 100.0 | 100.0 |

B) Atmospheric loss for 1 Earth mass planets with a 1 bar initial surface pressure

| Star RGB | Mars | Jupiter | Saturn | Kuiper Belt |
|---|---|---|---|---|
| 0.5 | * | * | * | 90.0 |
| 0.75 | * | * | * | 92.0 |
| 1 | * | 59.0 | 88.0 | 99.0 |
| 1.3 | * | 92.0 | 98.0 | 99.8 |
| 1.5 | 62.0 | 97.0 | 99.0 | 100.0 |
| 1.9 | 96.0 | 99.7 | 99.9 | 100.0 |

C) Atmospheric loss for super-Earths with 5 Earth masses with a 3.5 bar initial surface pressure

| Star RGB | Mars | Jupiter | Saturn | Kuiper Belt |
|---|---|---|---|---|
| 0.5 | * | * | 34 | 93.5 |
| 0.75 | * | * | 51.0 | 95.2 |
| 1 | * | 73.2 | 92.0 | 99.1 |
| 1.3 | 39.0 | 81.5 | 94.6 | 99.4 |
| 1.5 | 74.9 | 97.7 | 99.4 | 100.0 |
| 1.9 | 97.7 | 99.7 | 100.0 | 100.0 |

D) Atmospheric loss for super-Earths with 10 Earth masses with a 6.1 bar initial surface pressure

| Star RGB | Mars | Jupiter | Saturn | Kuiper Belt |
|---|---|---|---|---|
| 0.5 | * | * | 45 | 94.4 |
| 0.75 | * | * | 59.0 | 95.9 |
| 1 | * | 77.7 | 93.3 | 99.3 |
| 1.3 | 49.0 | 95.6 | 98.7 | 99.8 |
| 1.5 | 79.2 | 98.2 | 99.5 | 100.0 |
| 1.9 | 98.0 | 99.8 | 100.0 | 100.0 |

**TABLE IV**

Atmospheric mass loss during the AGB for the Sun – A5 grid stars (* denotes atmospheric surface pressure below 0.25 bar, where we terminate the calculations). All values are given in % of remaining atmosphere.

A) Atmospheric loss for super-moons with 0.5 Earth mass planets with a 0.57 bar initial surface pressure

| Star AGB | Mars | Jupiter | Saturn | Kuiper Belt |
|---|---|---|---|---|
| 1 | * | * | 72.5 | 97.3 |
| 1.3 | * | 67.5 | 90.3 | 98.9 |
| 1.5 | * | 78.6 | 93.5 | 99.3 |
| 1.9 | * | 90.2 | 97 | 99.6 |

B) Atmospheric loss for 1 Earth mass planets with a 1 bar initial surface pressure

| Star AGB | Mars | Jupiter | Saturn | Kuiper Belt |
|---|---|---|---|---|
| 1 | * | * | 77.3 | 97.8 |
| 1.3 | * | 73.2 | 92 | 99. |
| 1.5 | * | 82 | 95 | 99.3 |
| 1.9 | * | 92 | 98 | 100.0 |

C) Atmospheric loss for super-Earths with 5 Earth masses with a 3.5 bar initial surface pressure

| Star AGB | Mars | Jupiter | Saturn | Kuiper Belt |
|---|---|---|---|---|
| 1 | * | 49.9 | 85.3 | 98.8 |
| 1.3 | * | 82.6 | 93.3 | 99.5 |
| 1.5 | * | 88.6 | 96.8 | 99.7 |
| 1.9 | 38.3 | 94.9 | 98.7 | 100.0 |

D) Atmospheric loss for super-Earths with 10 Earth masses with a 6.1 bar initial surface pressure

| Star AGB | Mars | Jupiter | Saturn | Kuiper Belt |
|---|---|---|---|---|
| 1 | * | 58.1 | 87.5 | 98.7 |
| 1.3 | * | 86.7 | 96.1 | 99.7 |
| 1.5 | * | 91.5 | 97.5 | 99.8 |
| 1.9 | 57 | 96.2 | 98.9 | 1000 |

# APPENDIX B

We explore the sensitivity of the planetary atmospheric mass loss of Earth-mass planets in our solar system for different values of the entrainment efficiency ($\alpha$). Table IA shows results for a nominal $\alpha$ value of 0.2. Tables IB and IC show the results for both low (0.03) and high (0.3) values of $\alpha$ for comparison.

**Table I:**

A) Planetary mass loss assuming a nominal entrainment efficiency Alpha = 0.2

| Distance | Planet press (end RGB) | Planet Press (end AGB) |
|---|---|---|
| Mars | 0.25 | - |
| Jupiter | 0.59 | 0.25 |
| Saturn | 0.88 | 0.77 |
| Kuiper belt | 0.99 | 0.98 |

B) Planetary mass loss assuming a nominal entrainment efficiency Alpha = 0.03

| Distance | Planet press (end RGB) | Planet Press (end AGB) |
|---|---|---|
| Mars | 0.25 | - |
| Jupiter | 0.94 | 0.89 |
| Saturn | 0.98 | 0.97 |
| Kuiper belt | 1.00 | 1.00 |

C) Planetary mass loss assuming a nominal entrainment efficiency Alpha = 0.3

| Distance | Planet press (end RGB) | Planet press (end AGB) |
|---|---|---|
| Mars | 0.25 | - |
| Jupiter | 0.39 | 0.25 |
| Saturn | 0.82 | 0.66 |
| Kuiper belt | 0.98 | 0.97 |